\documentclass[fleqn,usenatbib]{mnras}

\usepackage{newtxtext,newtxmath}

\usepackage[T1]{fontenc}
\usepackage{ae,aecompl}

\usepackage{graphicx}	
\usepackage{amsmath}	
\usepackage{amssymb}	
\usepackage{color}

\title[GW transient rates, cosmic star formation and metallicity evolution]{Dependence of Gravitational Wave Transient Rates on Cosmic Star Formation and Metallicity Evolution History }

   \author[Tang, Eldridge, Stanway \& Bray]{Petra N. Tang$^1$
          \thanks{ntan032@aucklanduni.ac.nz},
          J. J. Eldridge$^1$\thanks{j.eldridge@auckland.ac.nz},
          Elizabeth R. Stanway$^2$ and
          J. C. Bray$^3$
          \\
          $^1$Department of Physics, University of Auckland, Private Bag 92019, Auckland, New Zealand\\
           $^2$Department of Physics, University of Warwick, Gibbet Hill Road, Coventry, CV4 7AL, UK\\
           $^3$Faculty of Science, Technology, Engineering and Mathematics, The Open University, Milton Keynes, UK
             }

\date{Accepted XXX. Received YYY; in original form ZZZ}

\pubyear{2019}

\begin{document}
\label{firstpage}
\pagerange{\pageref{firstpage}--\pageref{lastpage}}
\maketitle

\begin{abstract}
We compare the impacts of uncertainties in both binary population synthesis models and the cosmic star formation history on the predicted rates of Gravitational Wave compact binary merger (GW) events. These uncertainties cause the predicted rates of GW events to vary by up to an order of magnitude. 
Varying the volume-averaged star formation rate density history of the Universe causes the weakest change to our predictions, while varying the metallicity evolution has the strongest effect. Double neutron-star merger rates are more sensitive to assumed neutron-star kick velocity than the cosmic star formation history. Varying certain parameters affects merger rates in different ways depending on the mass of the merging compact objects; thus some of the degeneracy may be broken by looking at all the event rates rather than restricting ourselves to one class of mergers.
\end{abstract}




\section{Introduction}

Since the detection of the first confirmed gravitational wave compact binary merger events, astronomers and astrophysicists have been considering how to use this new window on the Universe to place constraints on its contents and our understanding of the underlying physics. \citep[e.g.][]{abbott2016improved,2017ApJ...848L..12A,2017Natur.551...85A,abbott2019gwtc}.
One of the most straightforward observables from GW events is their volumetric rate in the local Universe. A key test of stellar population synthesis codes is to reproduce that observed rate \citep[e.g.][]{2018MNRAS.477.4228P,2018MNRAS.481.1908K,eldridge2018consistent}. To do this population synthesis codes predict a delay-time distribution: the expected event rate of GW transients versus time for a given amount of star formation. This is then combined with an assumed star formation history, along with its metallicity evolution, to predict a rate at the current epoch \citep[see e.g.][]{langer2006collapsar,de2015merger,eldridge2018consistent}. There has been significant study of the how the uncertainties and assumptions in the stellar population models, especially the natal-supernova kick, affect the rate predictions and the observed double neutron star population \citep[e.g][]{1998ApJ...496..333F,2000ApJ...528..401W,2013ApJ...779...72D,2014MNRAS.440.1193L,abbott2016improved,2016MNRAS.456.4089B,2016ApJ...829L..13B,2017ApJ...846..170T,belczynski2017gw170104,2018MNRAS.481.1908K,2018MNRAS.481.4009V,2018MNRAS.474.2937C,2018MNRAS.479.4391M,eldridge2018consistent,2019MNRAS.482.2234G,2019ApJ...880L...8A}. However there are also substantial uncertainties in the star formation history of the Universe and its metallicity evolution which have been largely neglected until recently \citep{2018MNRAS.480.2704L,2019MNRAS.482.5012C,2019MNRAS.487....2M,2019MNRAS.487.1675A,2019arXiv190608136N}.

In this letter we build upon our initial work in \cite{2016MNRAS.462.3302E} and \cite{eldridge2018consistent} to investigate how varying the early star formation history and metallicity evolution of Universe affects the predicted event rate of GW transients. We compare this to the changes in the event rates from changing some of the parameters within our stellar population synthesis models. 

\section{Methods, Observations \& Simulations}

\begin{figure}
    \centering
    \includegraphics[width=\columnwidth]{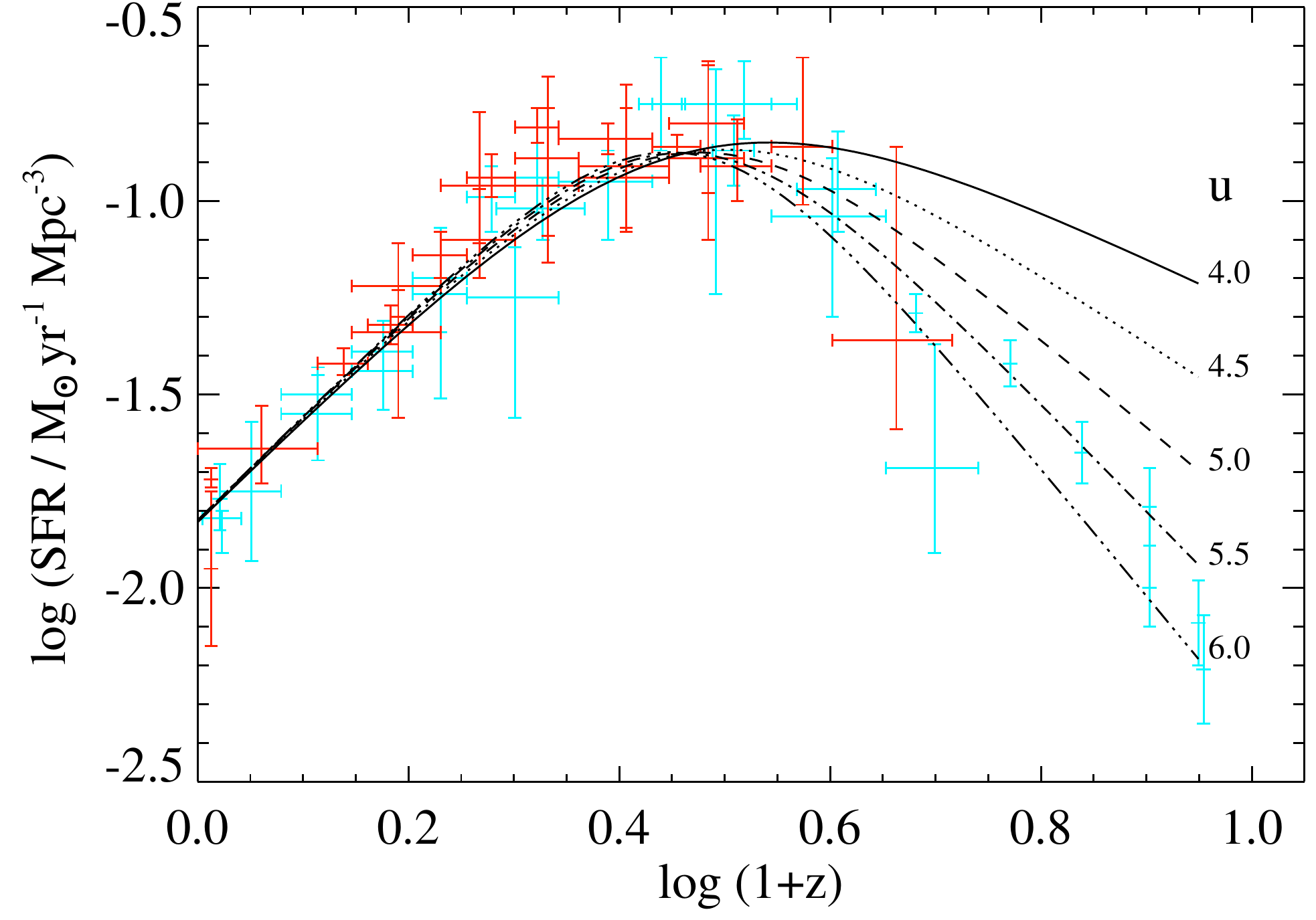}
    \vspace*{-20pt}
    \includegraphics[width=\columnwidth]{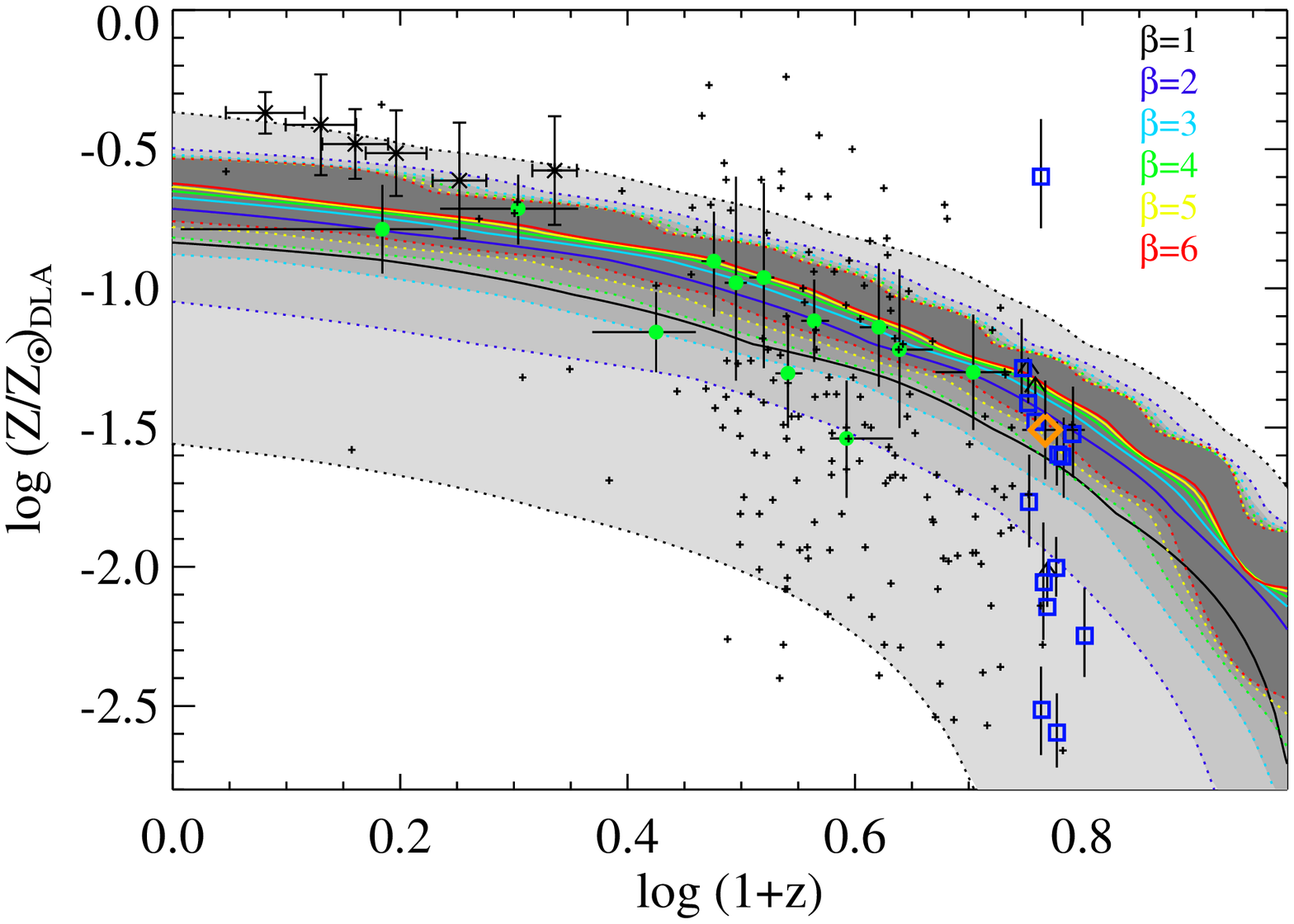}
    \caption{Comparison of our cosmic evolution models to observations. The upper panel shows the volume-averaged star formation rate density evolution, with observed values from \citep{madau2014cosmic}. Red points indicate estimates derived from infrared measurements, cyan points are derived from ultraviolet data. The lower panel shows our model for the median and 16-84th percentile range in metallicity as a function of redshift. These are compared to observed values derived from Damped Lyman-$\alpha$ (DLA) systems: green, blue and orange points from the compilation of \citet{2019MNRAS.tmp.2603P}, asterisks from \citet{2007A&A...462..429B}, small crosses from \citet{2012ApJ...755...89R}. Models are uniformly offset by -0.5 dex to account for the metallicity deficit of DLAs relative to typical star forming regions.}
    \label{fig:sfrhist}
\end{figure}

To calculate the GW transient event rate and its redshift evolution, we use the method outlined in \citet{eldridge2018consistent}. Delay-time distributions calculated from the Binary Population and Spectral Synthesis (BPASS\footnote{\url{http://bpass.auckland.ac.nz}}) v2.2.1 code \citep{2017PASA...34...58E,2018MNRAS.479...75S} were combined with a volume-averaged cosmic star formation history \citep{madau2014cosmic} and a model for the evolution of metallicity in star forming regions \citep{langer2006collapsar}. We now build upon this method to investigate how varying the star formation history and/or the cosmic metallicity evolution changes the expected rate of compact binary mergers. Specifically we consider neutron star-neutron star  (NS-NS), black hole-neutron star (BH-NS) and black hole-black hole (BH-BH) mergers.

We modify our method from \citet{eldridge2018consistent} in two ways. First, to allow for uncertainties in the high redshift cosmic star formation history we adopt a star formation rate density as a function of redshift as follows,
\begin{equation}
\rm \psi(z) = 0.015 \frac{(1+z)^{2.7}}{1+((1+z)/2.9)^{u}} \quad \rm M_{\odot}\,yr^{-1}\,Mpc^{-3}.
\end{equation}
This is a modification of the functional form given by \citet{madau2014cosmic}, where we have replaced the exponent in the denominator of 5.6 with a variable $u$. This alters how quickly the density of star formation in the early Universe increases, while having little impact on the density of star formation and the total stellar density observed today. Nonetheless, it could change the GW event rate significantly due to the expected long delay times of GW events. We allow $u$ to take values from 4 to 6, allowing for greater or lower star formation densities respectively at early times. Values of $u$ in the range for 5.5 to 6 give the best agreement with observations as shown in Figure \ref{fig:sfrhist}. However these data \citep[compiled from multiple sources by][]{madau2014cosmic} are derived from rest-frame ultraviolet measurements and so are subject to dust correction factors approaching 1\,dex. Galaxies at the highest redshifts are currently believed to have very little dust, based on fitting to their spectral energy distributions \citep{2012ApJ...754...83B} but the uncertainties on this are large \citep{2016MNRAS.455..659W,2018MNRAS.473.5363W}. If the intrinsic spectra of these sources are bluer (i.e. younger, lower metallicity or with a higher stellar rotation and/or multiple fraction) than currently estimated, the dust extinction will be underestimated, pushing the function towards lower $u$. We evaluate a large range of models to allow for this possibility.

In addition we calculate the fraction of star formation at different metallicities using the expression of \citet{langer2006collapsar}: 
\begin{equation}
\psi\left(\frac{Z}{Z_\odot}\right) = \frac{\hat{\Gamma}(\alpha+2,(Z/Z_\odot)^\beta)10^{0.15\beta z}}{\Gamma(\alpha+2)} \quad \rm M_{\odot} \, yr^{-1} \, Mpc^{-3}
\end{equation}
In this expression the value of $\beta$ determines how quickly the Universe becomes enriched with metals, $Z$, as a function of redshift $z$ and how broad the metallicity distribution is at each redshift. While in \citet{eldridge2018consistent} we used $\beta=2$ here we allow this exponent to vary from 1 to 6. A higher $\beta$ means the Universe was more quickly enriched, with a smaller metallicity scatter. In Figure \ref{fig:sfrhist} we compare the model metallicity enrichment for different $\beta$s to that determined from observations of Damped Lyman-$\alpha$ systems (DLAs). We offset the models by -0.5 dex to correct for the higher impact parameters (and thus lower measured metallicities) of DLAs relative to measurements in star forming galaxies \citep{2019arXiv190805362M}. The scatter in the data suggests that $\beta=1-2$ provides a better match to the observed spread than higher values, although the behaviour at the highest redshifts is largely unconstrained. If star formation occurs preferentially in already-enriched (i.e. more massive, older) dark matter halos in the distant Universe, rather than in sparse regions of the cosmic web, then a narrower range of metallicities might be expected for starbursts than is seen in the vast range of environments probed by DLAs. Again, we consider a broad range of $\beta$ values to allow for this possibility. 

Over this cosmic history parameter space we calculate three model grids with different BPASS stellar population synthesis model sets. These are as follows:
\begin{enumerate}
    \item BPASSv2.1 \citep{2017PASA...34...58E} models assume every star is in a binary with a flat distribution in mass ratio and the $\log$ of initial period. When a supernova occurs, a kick velocity is picked at random from the neutron star kick velocity distribution of \citet{2005MNRAS.360..974H}.
    \item BPASSv2.2, Hobbs \citep{2018MNRAS.479...75S} models use the empirical binary population and parameter distributions of \citet{2017ApJS..230...15M} and are typically more robust for the old stellar populations dominated by low mass stars. The supernova kick is also picked at random from the distribution of \citet{2005MNRAS.360..974H}.
    \item BPASSv2.2, Bray models use the same v2.2 initial binary parameter distributions as for (ii) but now we use the neutron star kick velocity from the work of \citet{2018MNRAS.480.5657B}:
    \begin{equation}
    v_{\rm kick2D}/\,{\rm km\,s^{-1}}= 100^{+30}_{-20}\,\left(\frac{M_{\rm ejecta}}{M_{\rm remnant}}\right) -170^{+100}_{-100}
\end{equation}
\end{enumerate}


We show the predicted $z=0$ GW event rates for these three BPASS model sets assuming the fiducial cosmic history parameters of $\beta=2$ and $u=5.6$ in Table \ref{tab:threemodels}. Varying the initial binary population, mass ratio and separation distributions has little effect on the NS-NS or NS-BH merger rate, but more than doubles the BH-BH merger rate. This is similar to results found by \citet{mandel2016merging} and \citet{belczynski2017gw170104}. However changing the SN kick model has a more general effect, with the Bray kick \citep{2018MNRAS.480.5657B} increasing the NS-NS and NS-BH rates while the BH-BH merger rate is decreased.  This indicates that the effects of these different stellar population synthesis assumptions on merger rates in different mass categories are orthogonal and that fitting all event categories at the same time will provide firmer constraints on the underlying physics of stellar models. 

The NS-NS and BH-BH merger rates from models (i) and (iii) are consistent with observations from LIGO Observing Run 2 (which are themselves still subject to significant uncertainty, see Table  \ref{tab:threemodels}), while in model (ii)  the BH-BH merger rate is too high. The observed NS-NS merger rate derived from the GW\,170817 event is higher than previously predicted \citep{2017PhRvL.119p1101A}. Thus while stellar population synthesis models (i) and (iii) are both consistent with the observed rates within the formal uncertainty, model (iii) is in best agreement with the data, assumping our fiducial star formation and metallicity histories. The next step, of course, is to vary this assumption.

\begin{table}
  \centering
  \caption{Event rates in Gpc$^{-3}$yr$^{-1}$, for varying stellar population synthesis assumptions and fiducial cosmic history model (u = 5.6, $\rm \beta$ = 2). Observational estimates based on the second LIGO observing run are given in the final column from \citet{2017PhRvL.119p1101A} for the NS-NS events and \citet{abbott2019binary}, their model B, for the BH-BH events. The uncertainties are the 90 per cent confidence limits.}
    \begin{tabular}{r|rrrr}
           events & v2.1, Hobbs & v2.2, Hobbs & v2.2, Bray & LIGO O2 \\
          \hline
    NS-NS & 472 & 417 & 2220 & $1504^{+3200}_{-1200}$\\
    NS-BH & 214 & 204 & 352 & -- \\
    BH-BH & 51.5 & 134 & 59.7 & $53.2^{+58.5}_{-28.8}$\\
    \end{tabular}%
  \label{tab:threemodels}%
\end{table}%

\section{Results}

\begin{figure*}
    \centering
    \includegraphics[width=1.8\columnwidth]{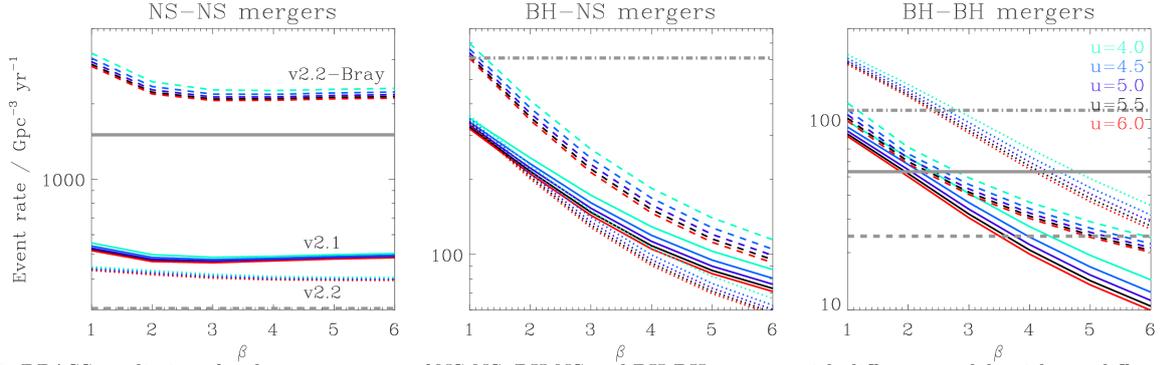}
    \vspace*{-15pt}
    \caption{BPASS predictions for the merger rates of NS-NS, BH-NS and BH-BH mergers with different models with our different models. The solid lines are for model (i) v2.1, the dotted-lines are for model (ii) v2.2 and the dashed-lines for model (iii) v2.2 with the Bray kick. The solid horizontal grey lines indicate the best estimate for the merger rates from \citet{abbott2016binary,abbott2019binary} while the dashed lines are the lower limits and the dash-dotted lines are the upper limits. For NS-NS, BH-NS and BH-BH mergers rates these are taken to be $1540^{+3200}_{-1200}$, $<$610 and $53.2^{+58.5}_{-28.8}$ $\rm Gpc^{-3}\, yr^{-1}$ respectively.    }
    
    \label{fig:combinedfigure}
\end{figure*}

We present the results of varying the early star formation rate parameter $u$ and the metallicity evolution parameter $\beta$ in Figure \ref{fig:combinedfigure} and provide quantitative values in Table \ref{tab:model3} in the appendix. The trends we observe are relatively simple.

Changing the early amount of star formation by varying $u$ has the weakest impact of our two parameters. Decreasing $u$ leads to greater early star formation and thus increases the number of low metallicity, massive early stars that can produce GW events after a long delay time. As a result the current-epoch rate of BH-BH and BH-NS events at fixed $\beta$ shows a weak tendency to increase at low values of $u$.

In comparison the metallicity evolution parameter $\beta$ has a much stronger effect, changing the merger rate by an order of magnitude for BS-NS and BH-BH mergers. The merger rates are highest at $\beta$ = 1, indicating that GW transients prefer a low metallicity environment. In all cases a higher $\beta$ leads to more metal-rich stars that  are typically less efficient at creating BH-BH and BH-NS GW events \citep[e.g][]{2010ApJ...715L.138B,2016MNRAS.462.3302E}. This is because higher metallicity stars have stronger stellar winds so lose more mass creating less massive black holes as well as wider binaries that have longer merger times via gravitational waves.

NS-NS mergers in comparison show much weaker dependence on $\beta$. At the highest $\beta$ values the merger rates reverse their decline and begin to increase instead, especially in models (i) and (iii). This is because as the stellar mass loss rates increase in high metallicity stars they become more likely to produce more neutron star than black holes at core collapse. The NS-NS merger rate increases at the expense of mergers involving black holes. We note that the almost flat dependence of the NS-NS merger rate on the star-formation parameters shows that this rate is more dependent on the stellar evolution model parameters than the star formation history. 

This confirms that the relative rates of different classes of events from LIGO/VIRGO Consortium observations will have diagnostic power in distinguishing between physically motivated models for both stellar physics effects and cosmic evolution histories.

On each of the panels we include lines representing the current observational constraints on rates \citep{abbott2016binary,abbott2019binary}. For the BH-BH mergers, models with $\beta$ close to 2 give values in good agreement with observations, suggesting that our fiducial model is a reasonable match to the observed Universe, although a range of $\beta$ from 1 to 3 are also consistent with the rate uncertainties.

The current observationally inferred NS-NS merger rate is significantly higher than most extant population synthesis codes predict \citep{2018MNRAS.474.2937C,2018MNRAS.479.4391M,2019MNRAS.482.5012C}. However, novel kick schemes or a different metallicity evolution history of the Universe have been shown to push predictions close to the observed constraint, with the highest being that from using the kick of \citet{2018MNRAS.480.5657B} as we see here. Both models (i) and (ii) are at the lower bound of observed rate while model (iii) is at the upper bound. This indicates that the assumed kick distribution may lie between these estimates and that GW event rates may prove an effective constraint on this distribution. Finally all cases with $\beta=2$ yield predictions in good agreement with  current upper limits on the NS-BH merger rate.

\section{Discussion \& Conclusions}

It is clear from our results that adopting different assumptions for the star formation and metallicity evolution of the Universe can have a significant effect on current epoch compact binary merger rates. These are as important, if not more so for the metallicity evolution, than changing the input physics of the binary interactions. We have found that the highest merger rates are obtained for low values of $\beta$ with little dependence on the values of $u$. However this primarily affects the merger rates involving black holes, the NS-NS merger rate depends only very weakly on the star formation history. Thus by using the different merger rates it may be possible to break the degeneracy between binary evolution and cosmic history uncertainties. We note that our results are in line with other recent work in highlighting the importance of the metallicity evolution in predicting the correct merger rates  \citep{2019MNRAS.482.5012C,2019arXiv190608136N}.

For model (i) (BPASS 2.1, Hobbs), varying $u$ and $\beta$ changes the predicted NS-NS rate from the fiducial value by $<10$ per cent in all cases except that of the lowest $\beta$ which reaches $18$ per cent at $u=4$. By contrast the NS-BH rate can vary by as much as 67 per cent from the fiducial value and the BH-BH rate can vary by $90$ per cent of the fiducial value. 

For model (ii) (BPASS 2.2, Hobbs), varying $u$ and $\beta$ changes the predicted NS-NS rate from the fiducial value by $<7$ per cent, the NS-BH rate by up to 77 per cent  and the BH-BH rate by up to $80$ per cent of the fiducial value. 

For model (iii) (BPASS 2.2, Bray), the pattern is similar with the predicted NS-NS rate varying from the fiducial value by $<10$ per cent across all parameters other than $\beta=1$ which can reach a 43 per cent variation at low $u$. The NS-BH and BH-BH rates both vary by a factor of 2 relative to the fiducial value over the parameter range explored.

In all cases, the variation seen (within a factor of two) suggests that the current observation of the NS-BH and BH-BH rates would struggle to discriminate between cosmic histories, given their substantial uncertainties. Measurements with a precision of a few percent will be required to do so, which may also lie beyond the capability of the current observing run (LVC O3).

By contrast, we have demonstrated that discriminating between the underlying assumptions of the stellar population synthesis models may be possible with weaker observational data. Changing the initial binary distribution (from v2.1 to v2.2) modestly reduces NS-NS and NS-BH events by $\sim$5-10 per cent but causes a dramatic, factor of 2.6, increase in the BH-BH rates due to the larger number of close binaries occurring in a stellar population of a given total initial mass using the updated prescription. More pronounced still is the effect of changing the supernova kick prescription, which boosts the rate of NS-NS mergers by almost an order of magnitude, since more low mass systems survive their two-supernova evolution pathways without being disassociated. Importantly, we notice that the impact of changing  cosmic history input parameters and physical model of stellar population synthesis vary by compact binary type. While interpreting event rates of one type (e.g. NS-NS) will lead to degenerate explanations, this degeneracy can be broken by comparing the rates of different types simulataneously. This implies that attempting to match the observed merger rate of these events at the same time will allow us to constrain our understanding of various aspects of stellar physics and also the evolution of the Universe.
 
A consistent picture from our results is that changing the assumed star formation history or metallicity evolution away from their current fiducial parameters generally leads to a decreased GW event rate. The predicted GW event rate only increases with an (unlikely) increase in the amount of stars formed in the early Universe. This would likely put the cosmic star formation rate density history in tension both with direct observations of the early universe \citep[e.g.][]{2019ApJ...880...25B} and with the observed history of mass assembly \citep[which should follow its integral, see][]{madau2014cosmic,2019MNRAS.tmp.2490W}. Alternately a higher rate may be possible with a much slower metallicity evolution that is at odds with the observed early enrichment of the Universe \citep[with relatively high metallicities being measured out to $z>3$, see e.g.][and Figure \ref{fig:sfrhist}]{2019arXiv190700013S}. The forthcoming James Webb Space Telescope will place much improved direct constraints on the properties of early star formation. In particular, it will improve observations of galaxies in the rest frame optical (with its relative insensitivity to dust extinction) and permit the true star formation rate density and metallicity at high redshift to be determined at unprecedented precision. While this work is already making progress from the ground at $z\sim2-3$ \citep[e.g.][]{2016ApJ...826..159S,2019MNRAS.tmp.2653S}, the sensitivity, wavelength coverage and multiplexing capability of NIRSPEC  will  permit determination of precise metallicities directly in the star forming galaxies which are responsible for the large delay-time tail of the GW event distributions. 

We note that predictions from BPASS stellar population synthesis models, when combined with the adopted fiducial SFH and metallicity evolution of u = 5.6 and $\rm \beta$ = 2 (from \citet{madau2014cosmic} and \citet{langer2006collapsar} respectively) are in good agreement with the current LIGO/Virgo BH-BH and NS-NS merger rate estimates. The high NS-NS rate derived from GW\,170817 favours model (iii), BPASS v2.2 with a Bray \& Eldridge (2018) SN kick, in particular. If the NS-NS merger rate derived in the current and future LIGO/VIRGO observing runs is indeed as high as we predict, this will provide further support for adoption of a revised neutron star kick distribution in future work. Combining predicted event rates with the chirp mass distribution will provide further constraints on stellar and cosmic evolution models, although care will be needed to account for the impact of chirp mass on detectability of systems when comparing models to data. 




\section*{Acknowledgements}

PNT acknowledges travel support from the University of Auckland. JJE acknowledges support from the University of Auckland and also the Royal Society Te Ap\={a}rangi of New Zealand under Marsden Fund. ERS acknowledges funding from the UK Science and Technology Research Council under grant ST/P000495/1.




\bibliographystyle{mnras}
\bibliography{references} 

\begin{thebibliography}{}
\makeatletter
\relax
\def\mn@urlcharsother{\let\do\@makeother \do\$\do\&\do\#\do\^\do\_\do\%\do\~}
\def\mn@doi{\begingroup\mn@urlcharsother \@ifnextchar [ {\mn@doi@}
  {\mn@doi@[]}}
\def\mn@doi@[#1]#2{\def\@tempa{#1}\ifx\@tempa\@empty \href
  {http://dx.doi.org/#2} {doi:#2}\else \href {http://dx.doi.org/#2} {#1}\fi
  \endgroup}
\def\mn@eprint#1#2{\mn@eprint@#1:#2::\@nil}
\def\mn@eprint@arXiv#1{\href {http://arxiv.org/abs/#1} {{\tt arXiv:#1}}}
\def\mn@eprint@dblp#1{\href {http://dblp.uni-trier.de/rec/bibtex/#1.xml}
  {dblp:#1}}
\def\mn@eprint@#1:#2:#3:#4\@nil{\def\@tempa {#1}\def\@tempb {#2}\def\@tempc
  {#3}\ifx \@tempc \@empty \let \@tempc \@tempb \let \@tempb \@tempa \fi \ifx
  \@tempb \@empty \def\@tempb {arXiv}\fi \@ifundefined
  {mn@eprint@\@tempb}{\@tempb:\@tempc}{\expandafter \expandafter \csname
  mn@eprint@\@tempb\endcsname \expandafter{\@tempc}}}

\bibitem[\protect\citeauthoryear{Abbott et~al.,}{Abbott
  et~al.}{2016a}]{abbott2016improved}
Abbott B.~P.,  et~al., 2016a, Physical Review X, 6, 041014

\bibitem[\protect\citeauthoryear{Abbott et~al.,}{Abbott
  et~al.}{2016b}]{abbott2016binary}
Abbott B.~P.,  et~al., 2016b, Physical Review X, 6, 041015

\bibitem[\protect\citeauthoryear{{Abbott} et~al.,}{{Abbott}
  et~al.}{2017a}]{2017PhRvL.119p1101A}
{Abbott} B.~P.,  et~al., 2017a, \mn@doi [\prl]
  {10.1103/PhysRevLett.119.161101}, \href
  {https://ui.adsabs.harvard.edu/abs/2017PhRvL.119p1101A} {119, 161101}

\bibitem[\protect\citeauthoryear{{Abbott} et~al.,}{{Abbott}
  et~al.}{2017b}]{2017Natur.551...85A}
{Abbott} B.~P.,  et~al., 2017b, \mn@doi [\nat] {10.1038/nature24471}, \href
  {https://ui.adsabs.harvard.edu/abs/2017Natur.551...85A} {551, 85}

\bibitem[\protect\citeauthoryear{{Abbott} et~al.,}{{Abbott}
  et~al.}{2017c}]{2017ApJ...848L..12A}
{Abbott} B.~P.,  et~al., 2017c, \mn@doi [\apjl] {10.3847/2041-8213/aa91c9},
  \href {https://ui.adsabs.harvard.edu/abs/2017ApJ...848L..12A} {848, L12}

\bibitem[\protect\citeauthoryear{Abbott et~al.,}{Abbott
  et~al.}{2019a}]{abbott2019binary}
Abbott B.,  et~al., 2019a, arXiv preprint arXiv:1811.12940

\bibitem[\protect\citeauthoryear{Abbott et~al.,}{Abbott
  et~al.}{2019b}]{abbott2019gwtc}
Abbott B.,  et~al., 2019b, Physical Review X, 9, 031040

\bibitem[\protect\citeauthoryear{{Andrews} \& {Mandel}}{{Andrews} \&
  {Mandel}}{2019}]{2019ApJ...880L...8A}
{Andrews} J.~J.,  {Mandel} I.,  2019, \mn@doi [\apjl]
  {10.3847/2041-8213/ab2ed1}, \href
  {https://ui.adsabs.harvard.edu/abs/2019ApJ...880L...8A} {880, L8}

\bibitem[\protect\citeauthoryear{{Artale}, {Mapelli}, {Giacobbo}, {Sabha},
  {Spera}, {Santoliquido}  \& {Bressan}}{{Artale}
  et~al.}{2019}]{2019MNRAS.487.1675A}
{Artale} M.~C.,  {Mapelli} M.,  {Giacobbo} N.,  {Sabha} N.~B.,  {Spera} M.,
  {Santoliquido} F.,   {Bressan} A.,  2019, \mn@doi [\mnras]
  {10.1093/mnras/stz1382}, \href
  {https://ui.adsabs.harvard.edu/abs/2019MNRAS.487.1675A} {487, 1675}

\bibitem[\protect\citeauthoryear{{Balestra}, {Tozzi}, {Ettori}, {Rosati},
  {Borgani}, {Mainieri}, {Norman}  \& {Viola}}{{Balestra}
  et~al.}{2007}]{2007A&A...462..429B}
{Balestra} I.,  {Tozzi} P.,  {Ettori} S.,  {Rosati} P.,  {Borgani} S.,
  {Mainieri} V.,  {Norman} C.,   {Viola} M.,  2007, \mn@doi [\aap]
  {10.1051/0004-6361:20065568}, \href
  {https://ui.adsabs.harvard.edu/abs/2007A&A...462..429B} {462, 429}

\bibitem[\protect\citeauthoryear{{Belczynski}, {Dominik}, {Bulik},
  {O'Shaughnessy}, {Fryer}  \& {Holz}}{{Belczynski}
  et~al.}{2010}]{2010ApJ...715L.138B}
{Belczynski} K.,  {Dominik} M.,  {Bulik} T.,  {O'Shaughnessy} R.,  {Fryer} C.,
   {Holz} D.~E.,  2010, \mn@doi [\apjl] {10.1088/2041-8205/715/2/L138}, \href
  {https://ui.adsabs.harvard.edu/abs/2010ApJ...715L.138B} {715, L138}

\bibitem[\protect\citeauthoryear{Belczynski et~al.,}{Belczynski
  et~al.}{2017}]{belczynski2017gw170104}
Belczynski K.,  et~al., 2017, preprint, 2, 2

\bibitem[\protect\citeauthoryear{{Beniamini} \& {Piran}}{{Beniamini} \&
  {Piran}}{2016}]{2016MNRAS.456.4089B}
{Beniamini} P.,  {Piran} T.,  2016, \mn@doi [\mnras] {10.1093/mnras/stv2903},
  \href {https://ui.adsabs.harvard.edu/abs/2016MNRAS.456.4089B} {456, 4089}

\bibitem[\protect\citeauthoryear{{Beniamini}, {Hotokezaka}  \&
  {Piran}}{{Beniamini} et~al.}{2016}]{2016ApJ...829L..13B}
{Beniamini} P.,  {Hotokezaka} K.,   {Piran} T.,  2016, \mn@doi [\apjl]
  {10.3847/2041-8205/829/1/L13}, \href
  {https://ui.adsabs.harvard.edu/abs/2016ApJ...829L..13B} {829, L13}

\bibitem[\protect\citeauthoryear{{Bouwens} et~al.,}{{Bouwens}
  et~al.}{2012}]{2012ApJ...754...83B}
{Bouwens} R.~J.,  et~al., 2012, \mn@doi [\apj] {10.1088/0004-637X/754/2/83},
  \href {https://ui.adsabs.harvard.edu/abs/2012ApJ...754...83B} {754, 83}

\bibitem[\protect\citeauthoryear{{Bouwens}, {Stefanon}, {Oesch}, {Illingworth},
  {Nanayakkara}, {Roberts-Borsani}, {Labb{\'e}}  \& {Smit}}{{Bouwens}
  et~al.}{2019}]{2019ApJ...880...25B}
{Bouwens} R.~J.,  {Stefanon} M.,  {Oesch} P.~A.,  {Illingworth} G.~D.,
  {Nanayakkara} T.,  {Roberts-Borsani} G.,  {Labb{\'e}} I.,   {Smit} R.,  2019,
  \mn@doi [\apj] {10.3847/1538-4357/ab24c5}, \href
  {https://ui.adsabs.harvard.edu/abs/2019ApJ...880...25B} {880, 25}

\bibitem[\protect\citeauthoryear{{Bray} \& {Eldridge}}{{Bray} \&
  {Eldridge}}{2018}]{2018MNRAS.480.5657B}
{Bray} J.~C.,  {Eldridge} J.~J.,  2018, \mn@doi [\mnras]
  {10.1093/mnras/sty2230}, \href
  {https://ui.adsabs.harvard.edu/abs/2018MNRAS.480.5657B} {480, 5657}

\bibitem[\protect\citeauthoryear{{Chruslinska}, {Belczynski}, {Klencki}  \&
  {Benacquista}}{{Chruslinska} et~al.}{2018}]{2018MNRAS.474.2937C}
{Chruslinska} M.,  {Belczynski} K.,  {Klencki} J.,   {Benacquista} M.,  2018,
  \mn@doi [\mnras] {10.1093/mnras/stx2923}, \href
  {https://ui.adsabs.harvard.edu/abs/2018MNRAS.474.2937C} {474, 2937}

\bibitem[\protect\citeauthoryear{{Chruslinska}, {Nelemans}  \&
  {Belczynski}}{{Chruslinska} et~al.}{2019}]{2019MNRAS.482.5012C}
{Chruslinska} M.,  {Nelemans} G.,   {Belczynski} K.,  2019, \mn@doi [\mnras]
  {10.1093/mnras/sty3087}, \href
  {https://ui.adsabs.harvard.edu/abs/2019MNRAS.482.5012C} {482, 5012}

\bibitem[\protect\citeauthoryear{{Dominik}, {Belczynski}, {Fryer}, {Holz},
  {Berti}, {Bulik}, {Mand el}  \& {O'Shaughnessy}}{{Dominik}
  et~al.}{2013}]{2013ApJ...779...72D}
{Dominik} M.,  {Belczynski} K.,  {Fryer} C.,  {Holz} D.~E.,  {Berti} E.,
  {Bulik} T.,  {Mand el} I.,   {O'Shaughnessy} R.,  2013, \mn@doi [\apj]
  {10.1088/0004-637X/779/1/72}, \href
  {https://ui.adsabs.harvard.edu/abs/2013ApJ...779...72D} {779, 72}

\bibitem[\protect\citeauthoryear{{Eldridge} \& {Stanway}}{{Eldridge} \&
  {Stanway}}{2016}]{2016MNRAS.462.3302E}
{Eldridge} J.~J.,  {Stanway} E.~R.,  2016, \mn@doi [\mnras]
  {10.1093/mnras/stw1772}, \href
  {https://ui.adsabs.harvard.edu/abs/2016MNRAS.462.3302E} {462, 3302}

\bibitem[\protect\citeauthoryear{{Eldridge}, {Stanway}, {Xiao}, {McClelland },
  {Taylor}, {Ng}, {Greis}  \& {Bray}}{{Eldridge}
  et~al.}{2017}]{2017PASA...34...58E}
{Eldridge} J.~J.,  {Stanway} E.~R.,  {Xiao} L.,  {McClelland } L.~A.~S.,
  {Taylor} G.,  {Ng} M.,  {Greis} S.~M.~L.,   {Bray} J.~C.,  2017, \mn@doi
  [\pasa] {10.1017/pasa.2017.51}, \href
  {https://ui.adsabs.harvard.edu/abs/2017PASA...34...58E} {34, e058}

\bibitem[\protect\citeauthoryear{{Eldridge}, {Stanway}  \& {Tang}}{{Eldridge}
  et~al.}{2019}]{eldridge2018consistent}
{Eldridge} J.~J.,  {Stanway} E.~R.,   {Tang} P.~N.,  2019, \mn@doi [\mnras]
  {10.1093/mnras/sty2714}, \href
  {https://ui.adsabs.harvard.edu/abs/2019MNRAS.482..870E} {482, 870}

\bibitem[\protect\citeauthoryear{{Fryer}, {Burrows}  \& {Benz}}{{Fryer}
  et~al.}{1998}]{1998ApJ...496..333F}
{Fryer} C.,  {Burrows} A.,   {Benz} W.,  1998, \mn@doi [\apj] {10.1086/305348},
  \href {https://ui.adsabs.harvard.edu/abs/1998ApJ...496..333F} {496, 333}

\bibitem[\protect\citeauthoryear{{Giacobbo} \& {Mapelli}}{{Giacobbo} \&
  {Mapelli}}{2019}]{2019MNRAS.482.2234G}
{Giacobbo} N.,  {Mapelli} M.,  2019, \mn@doi [\mnras] {10.1093/mnras/sty2848},
  \href {https://ui.adsabs.harvard.edu/abs/2019MNRAS.482.2234G} {482, 2234}

\bibitem[\protect\citeauthoryear{{Hobbs}, {Lorimer}, {Lyne}  \&
  {Kramer}}{{Hobbs} et~al.}{2005}]{2005MNRAS.360..974H}
{Hobbs} G.,  {Lorimer} D.~R.,  {Lyne} A.~G.,   {Kramer} M.,  2005, \mn@doi
  [\mnras] {10.1111/j.1365-2966.2005.09087.x}, \href
  {https://ui.adsabs.harvard.edu/abs/2005MNRAS.360..974H} {360, 974}

\bibitem[\protect\citeauthoryear{{Kruckow}, {Tauris}, {Langer}, {Kramer}  \&
  {Izzard}}{{Kruckow} et~al.}{2018}]{2018MNRAS.481.1908K}
{Kruckow} M.~U.,  {Tauris} T.~M.,  {Langer} N.,  {Kramer} M.,   {Izzard} R.~G.,
   2018, \mn@doi [\mnras] {10.1093/mnras/sty2190}, \href
  {https://ui.adsabs.harvard.edu/abs/2018MNRAS.481.1908K} {481, 1908}

\bibitem[\protect\citeauthoryear{{Lamberts} et~al.,}{{Lamberts}
  et~al.}{2018}]{2018MNRAS.480.2704L}
{Lamberts} A.,  et~al., 2018, \mn@doi [\mnras] {10.1093/mnras/sty2035}, \href
  {https://ui.adsabs.harvard.edu/abs/2018MNRAS.480.2704L} {480, 2704}

\bibitem[\protect\citeauthoryear{Langer \& Norman}{Langer \&
  Norman}{2006}]{langer2006collapsar}
Langer N.,  Norman C.,  2006, The Astrophysical Journal Letters, 638, L63

\bibitem[\protect\citeauthoryear{{Lipunov} \& {Pruzhinskaya}}{{Lipunov} \&
  {Pruzhinskaya}}{2014}]{2014MNRAS.440.1193L}
{Lipunov} V.~M.,  {Pruzhinskaya} M.~V.,  2014, \mn@doi [\mnras]
  {10.1093/mnras/stu313}, \href
  {https://ui.adsabs.harvard.edu/abs/2014MNRAS.440.1193L} {440, 1193}

\bibitem[\protect\citeauthoryear{Madau \& Dickinson}{Madau \&
  Dickinson}{2014}]{madau2014cosmic}
Madau P.,  Dickinson M.,  2014, Annual Review of Astronomy and Astrophysics,
  52, 415

\bibitem[\protect\citeauthoryear{{Mandel} \& {de Mink}}{{Mandel} \& {de
  Mink}}{2016}]{mandel2016merging}
{Mandel} I.,  {de Mink} S.~E.,  2016, \mn@doi [\mnras] {10.1093/mnras/stw379},
  \href {https://ui.adsabs.harvard.edu/abs/2016MNRAS.458.2634M} {458, 2634}

\bibitem[\protect\citeauthoryear{{Mapelli} \& {Giacobbo}}{{Mapelli} \&
  {Giacobbo}}{2018}]{2018MNRAS.479.4391M}
{Mapelli} M.,  {Giacobbo} N.,  2018, \mn@doi [\mnras] {10.1093/mnras/sty1613},
  \href {https://ui.adsabs.harvard.edu/abs/2018MNRAS.479.4391M} {479, 4391}

\bibitem[\protect\citeauthoryear{{Mapelli}, {Giacobbo}, {Santoliquido}  \&
  {Artale}}{{Mapelli} et~al.}{2019}]{2019MNRAS.487....2M}
{Mapelli} M.,  {Giacobbo} N.,  {Santoliquido} F.,   {Artale} M.~C.,  2019,
  \mn@doi [\mnras] {10.1093/mnras/stz1150}, \href
  {https://ui.adsabs.harvard.edu/abs/2019MNRAS.487....2M} {487, 2}

\bibitem[\protect\citeauthoryear{{Moe} \& {Di Stefano}}{{Moe} \& {Di
  Stefano}}{2017}]{2017ApJS..230...15M}
{Moe} M.,  {Di Stefano} R.,  2017, \mn@doi [\apjs] {10.3847/1538-4365/aa6fb6},
  \href {https://ui.adsabs.harvard.edu/abs/2017ApJS..230...15M} {230, 15}

\bibitem[\protect\citeauthoryear{{M{\o}ller} \& {Christensen}}{{M{\o}ller} \&
  {Christensen}}{2019}]{2019arXiv190805362M}
{M{\o}ller} P.,  {Christensen} L.,  2019, arXiv e-prints, \href
  {https://ui.adsabs.harvard.edu/abs/2019arXiv190805362M} {p. arXiv:1908.05362}

\bibitem[\protect\citeauthoryear{{Neijssel} et~al.,}{{Neijssel}
  et~al.}{2019}]{2019arXiv190608136N}
{Neijssel} C.~J.,  et~al., 2019, arXiv e-prints, \href
  {https://ui.adsabs.harvard.edu/abs/2019arXiv190608136N} {p. arXiv:1906.08136}

\bibitem[\protect\citeauthoryear{{Perna}, {Chruslinska}, {Corsi}  \&
  {Belczynski}}{{Perna} et~al.}{2018}]{2018MNRAS.477.4228P}
{Perna} R.,  {Chruslinska} M.,  {Corsi} A.,   {Belczynski} K.,  2018, \mn@doi
  [\mnras] {10.1093/mnras/sty814}, \href
  {https://ui.adsabs.harvard.edu/abs/2018MNRAS.477.4228P} {477, 4228}

\bibitem[\protect\citeauthoryear{{Poudel}, {Kulkarni}, {Cashman}, {Frye},
  {P{\'e}roux}, {Rahmani}  \& {Quiret}}{{Poudel}
  et~al.}{2019}]{2019MNRAS.tmp.2603P}
{Poudel} S.,  {Kulkarni} V.~P.,  {Cashman} F.~H.,  {Frye} B.,  {P{\'e}roux} C.,
   {Rahmani} H.,   {Quiret} S.,  2019, \mn@doi [\mnras]
  {10.1093/mnras/stz3000}, \href
  {https://ui.adsabs.harvard.edu/abs/2019MNRAS.tmp.2603P} {p.~2603}

\bibitem[\protect\citeauthoryear{{Rafelski}, {Wolfe}, {Prochaska}, {Neeleman}
  \& {Mendez}}{{Rafelski} et~al.}{2012}]{2012ApJ...755...89R}
{Rafelski} M.,  {Wolfe} A.~M.,  {Prochaska} J.~X.,  {Neeleman} M.,   {Mendez}
  A.~J.,  2012, \mn@doi [\apj] {10.1088/0004-637X/755/2/89}, \href
  {https://ui.adsabs.harvard.edu/abs/2012ApJ...755...89R} {755, 89}

\bibitem[\protect\citeauthoryear{{Sanders} et~al.,}{{Sanders}
  et~al.}{2019a}]{2019MNRAS.tmp.2653S}
{Sanders} R.~L.,  et~al., 2019a, \mn@doi [\mnras] {10.1093/mnras/stz3032},
  \href {https://ui.adsabs.harvard.edu/abs/2019MNRAS.tmp.2653S} {p.~2653}

\bibitem[\protect\citeauthoryear{{Sanders} et~al.,}{{Sanders}
  et~al.}{2019b}]{2019arXiv190700013S}
{Sanders} R.~L.,  et~al., 2019b, arXiv e-prints, \href
  {https://ui.adsabs.harvard.edu/abs/2019arXiv190700013S} {p. arXiv:1907.00013}

\bibitem[\protect\citeauthoryear{{Stanway} \& {Eldridge}}{{Stanway} \&
  {Eldridge}}{2018}]{2018MNRAS.479...75S}
{Stanway} E.~R.,  {Eldridge} J.~J.,  2018, \mn@doi [\mnras]
  {10.1093/mnras/sty1353}, \href
  {https://ui.adsabs.harvard.edu/abs/2018MNRAS.479...75S} {479, 75}

\bibitem[\protect\citeauthoryear{{Steidel}, {Strom}, {Pettini}, {Rudie},
  {Reddy}  \& {Trainor}}{{Steidel} et~al.}{2016}]{2016ApJ...826..159S}
{Steidel} C.~C.,  {Strom} A.~L.,  {Pettini} M.,  {Rudie} G.~C.,  {Reddy} N.~A.,
    {Trainor} R.~F.,  2016, \mn@doi [\apj] {10.3847/0004-637X/826/2/159}, \href
  {https://ui.adsabs.harvard.edu/abs/2016ApJ...826..159S} {826, 159}

\bibitem[\protect\citeauthoryear{{Tauris} et~al.,}{{Tauris}
  et~al.}{2017}]{2017ApJ...846..170T}
{Tauris} T.~M.,  et~al., 2017, \mn@doi [\apj] {10.3847/1538-4357/aa7e89}, \href
  {https://ui.adsabs.harvard.edu/abs/2017ApJ...846..170T} {846, 170}

\bibitem[\protect\citeauthoryear{{Vigna-G{\'o}mez} et~al.,}{{Vigna-G{\'o}mez}
  et~al.}{2018}]{2018MNRAS.481.4009V}
{Vigna-G{\'o}mez} A.,  et~al., 2018, \mn@doi [\mnras] {10.1093/mnras/sty2463},
  \href {https://ui.adsabs.harvard.edu/abs/2018MNRAS.481.4009V} {481, 4009}

\bibitem[\protect\citeauthoryear{{Wex}, {Kalogera}  \& {Kramer}}{{Wex}
  et~al.}{2000}]{2000ApJ...528..401W}
{Wex} N.,  {Kalogera} V.,   {Kramer} M.,  2000, \mn@doi [\apj]
  {10.1086/308148}, \href
  {https://ui.adsabs.harvard.edu/abs/2000ApJ...528..401W} {528, 401}

\bibitem[\protect\citeauthoryear{{Wilkins}, {Bouwens}, {Oesch}, {Labb{\'e}},
  {Sargent}, {Caruana}, {Wardlow}  \& {Clay}}{{Wilkins}
  et~al.}{2016}]{2016MNRAS.455..659W}
{Wilkins} S.~M.,  {Bouwens} R.~J.,  {Oesch} P.~A.,  {Labb{\'e}} I.,  {Sargent}
  M.,  {Caruana} J.,  {Wardlow} J.,   {Clay} S.,  2016, \mn@doi [\mnras]
  {10.1093/mnras/stv2263}, \href
  {https://ui.adsabs.harvard.edu/abs/2016MNRAS.455..659W} {455, 659}

\bibitem[\protect\citeauthoryear{{Wilkins}, {Feng}, {Di Matteo}, {Croft},
  {Lovell}  \& {Thomas}}{{Wilkins} et~al.}{2018}]{2018MNRAS.473.5363W}
{Wilkins} S.~M.,  {Feng} Y.,  {Di Matteo} T.,  {Croft} R.,  {Lovell} C.~C.,
  {Thomas} P.,  2018, \mn@doi [\mnras] {10.1093/mnras/stx2588}, \href
  {https://ui.adsabs.harvard.edu/abs/2018MNRAS.473.5363W} {473, 5363}

\bibitem[\protect\citeauthoryear{{Wilkins}, {Lovell}  \& {Stanway}}{{Wilkins}
  et~al.}{2019}]{2019MNRAS.tmp.2490W}
{Wilkins} S.~M.,  {Lovell} C.~C.,   {Stanway} E.~R.,  2019, \mn@doi [\mnras]
  {10.1093/mnras/stz2894}, \href
  {https://ui.adsabs.harvard.edu/abs/2019MNRAS.tmp.2490W} {p.~2490}

\bibitem[\protect\citeauthoryear{de Mink \& Belczynski}{de~Mink \&
  Belczynski}{2015}]{de2015merger}
de Mink S.,  Belczynski K.,  2015, The Astrophysical Journal, 814, 58

\makeatother
\end{thebibliography}




\appendix

\section{Tabulated results of GW event rates}

\begin{table}
  \centering
  \caption{Gravitational Wave event rates in Gpc$^{-3}$ yr$^{-1}$ to 3 significant figures.}
    \begin{tabular}{ccccccc}
    \hline
    &$\beta$ & \multicolumn{1}{l}{$u = 4$} & \multicolumn{1}{l}{$u = 4.5$} & \multicolumn{1}{l}{$u = 5$} & \multicolumn{1}{l}{$u = 5.5$} & \multicolumn{1}{l}{$u = 6$}\\
    \hline
 NS-NS &      1	& 556 & 541 & 531 & 523 & 518\\
model (i)&    2 & 499 & 487 & 479 & 473 & 468\\
&     3 & 487 & 477 & 471 & 467 & 463\\
&     4 & 491 & 483 & 478 & 474 & 471\\
&     5 & 499 & 491 & 486 & 482 & 480\\
&     6 & 504 & 497 & 492 & 488 & 485\\
    \hline
NS-BH&     1 & 350 & 338 & 330 &	323 & 318\\
model (i)&     2 & 243 & 230 & 222 & 215 & 210\\
&     3 & 173 & 161 & 153 & 148 & 143\\
&     4 & 129 & 119 & 113 & 108 & 105\\
&   5  & 103 & 95.0 & 89.8 & 86.3 & 83.8\\
&  6  & 87.1 & 80.3 & 76.0 & 73.2 & 71.1\\
\hline
BH-BH& 1 & 97.6 & 91.2 & 86.8 & 83.8 & 81.5\\
model (i)&     2 & 63.2 & 58.0 & 54.5 & 52.0 & 50.1\\
&    3 & 40.8 & 36.6 & 33.8 & 31.9 & 30.5\\
&    4 & 27.4 & 24.1 & 22.1 & 20.6 & 19.7\\
&    5 & 19.4 & 16.9 & 15.3 & 14.3 & 13.6\\
&    6 & 14.4 & 12.4 & 11.2 & 10.5 & 9.97\\
       \hline
 NS-NS&     1	    & 447 & 441 & 437 & 434 & 432\\
model (ii) &     2     & 431 & 425 & 421 & 418 & 415\\
  &    3     & 417 & 411 & 407 & 404 & 402\\
  &    4     & 409 & 404 & 400 & 398 & 396\\
     & 5     & 406 & 401 & 398 & 396 & 394\\
     & 6     & 405 & 401 & 398 & 396 & 394\\
    \hline
    NS-BH &        1	  & 360 & 349 &	342 & 336 & 332\\
 model (ii)    & 2     & 227 & 217 & 210 & 205 & 201\\
     & 3     & 149 & 141 & 135 & 131 & 128\\
     & 4     & 106 & 99.6 & 95.3 & 92.4 & 90.4\\
     & 5     & 81.7 & 76.6 & 73.4 & 71.3 & 69.8\\
     & 6     & 66.6 & 62.6 & 60.2 & 58.6 & 57.5\\
    \hline
 BH-BH &         1	  & 219& 209 &	203 & 198 &	194\\
model (ii)     & 2     & 153 & 145 & 139 & 135 & 132\\
     & 3     & 104 & 96.3 & 91.2 & 87.5 & 84.8\\
     & 4     & 70.2 & 63.9 & 59.8 & 56.9 & 54.8\\
     & 5     & 48.9 & 43.9 & 40.7 & 38.5 & 37.1\\
     & 6     & 35.4 & 31.5 & 29.1 & 27.6 & 26.5\\
    \hline
NS-NS&      1	    &3190 & 3050 & 2950 & 2880& 2830\\			
model (iii)     & 2     & 2450 & 2350 & 2280 & 2230 & 2190 \\
     & 3     & 2270 & 2190 & 2130 & 2090 & 2060 \\
     & 4     & 2260 & 2180 & 2130 & 2090 & 2070 \\
     & 5     & 2290 & 2210 & 2160 & 2120 & 2100 \\
     & 6     & 2310 & 2230 & 2180 & 2140 & 2110 \\
    \hline
  NS-BH &       1	    & 698 & 664 & 639& 6225 & 609\\			
 model (iii)    & 2     & 413 & 385 & 367 & 354 & 344\\
     & 3     & 264 & 243 & 229 & 220 & 213\\
     & 4     & 184 & 168 & 158 & 151 & 147\\
     & 5     & 140 & 128 & 120 & 115 & 112\\
     & 6     & 115 & 105 & 99.1 & 95.3 & 92.6\\
    \hline
   BH-BH &       1	    & 122  & 112 &	106  & 101 &  98.2\\			
model (iii)     & 2     & 71.4 & 66.0 & 62.4 & 60.1 & 58.4\\
     & 3     & 49.4 & 45.6 & 43.2 & 41.5 & 40.4\\
     & 4     & 36.8 & 34.0 & 32.3 & 31.1 & 30.2\\
     & 5     & 29.1 & 26.9 & 25.6 & 24.7 & 24.1\\
     & 6     & 24.0 & 22.2 & 21.2 & 20.5 & 20.1\\
     \hline
    \end{tabular}%
 \label{tab:model3}%
\end{table}%

\bsp	
\label{lastpage}
\end{document}